\documentclass[12pt]{iopart}
\usepackage[dvips]{graphicx} %LaTeX2e

\usepackage[round,authoryear]{natbib}
\bibpunct{(}{)}{,}{a}{}{,}

\begin{document}

\title[Probabilistic two-stage model of cell inactivation by ionizing particles]{Probabilistic two-stage model of cell inactivation by ionizing particles}

\author{Pavel Kundr\'{a}t, Milo\v{s} Lokaj\'{\i}\v{c}ek, Hana Hrom\v{c}\'{\i}kov\'{a}}

\address{Institute~of~Physics, Academy~of~Sciences~of~the~Czech~Republic,\\
Na~Slovance~2, CZ-182~21~Praha~8, Czech~Republic\\
E-mail: Pavel.Kundrat@fzu.cz}

\begin{abstract}
Model of biological effects of ionizing particles, especially of protons and other ions, is proposed. The model is based on distinguishing the single-particle and collective effects of the underlying radiobiological mechanism. The probabilities of individual particles to form severe damages to DNA, their synergetic or saturation combinations, and the effect of cellular repair system are taken into account. The model enables to describe linear, parabolic and more complex curves, including those exhibiting low-dose hypersensitivity phenomena, in a systematic way. Global shape as well as detailed structure of survival curves might be represented, which is crucial if different fractionation schemes in radiotherapy should be assessed precisely. Experimental cell-survival data for inactivation of V79 cells by low-energy protons have been analyzed and corresponding detailed characteristics of the inactivation mechanism have been derived for this case.
\end{abstract}



\maketitle

%%%%%%%%%%%%%%%%%%%%%%%%%%%%%%%%%%%%%%%%%%%%%%%%%%%%%%%%%%%%%%%%%%%%%%%%%%%%%%%%%%%%
\section{Introduction}
\label{sec:Introduction}
Models estimating the fraction of surviving cells are of utmost importance for applications of ionizing radiation in radiotherapy, because both tumor control and normal tissue complication probabilities (TCP, NTCP) are given primarily by cell survival in the corresponding regions. In the case of hadron radiotherapy, based on irradiating by accelerated protons and light ions, the biological effects are not given by applied dose only, but depend significantly on other physical characteristics of these particles. Their relative biological effectiveness (RBE) has been shown to vary hugely with ion kind and energy or LET value; see, e.g., \citep{Belli, Belli 2000, Folkard, Weyrather}.

Several phenomenological model approaches aimed at estimating the fraction of surviving cells have been proposed so far for treatment planning procedures, based e.g.\ on a similarity in LET characteristics between carbon and neutron beams \citep{HIMAC} or using the linear-quadratic (LQ) formalism with its $\alpha$ and $\beta$ coefficients depending on LET and ion kind \citep{Belli-Semiempirical}. A systematic model approach has been represented by the local effect model (LEM) of Scholz and Kraft \citep{LEM-1, Scholz+Kraft-1995, LEM-2, Scholz 1999}, in which the differences in biological effects of diverse ions have been related to their track structures. However, in any of these approaches the underlying physically-chemical and especially biological processes have not been addressed in detail. A realistic model scheme ("the probabilistic two-stage model") based on probabilistic description of the main processes involved in the corresponding radiobiological mechanism will be presented in this paper.

The probabilistic two-stage model is based, in principle, on distinguishing the single-particle and collective effects of the radiobiological mechanism: Individual events of energy transfer, subsequent chemical reactions and also formation of DNA damages occur practically independently after the impacts of individual ionizing particles, at least at dose levels and dose rates that are used commonly in radiotherapy. On the other hand, the final response of a given cell, which includes biological processes that start from DNA repair and lead finally to cell survival or inactivation, has to be classified as a response to the total effect of all traversing particles.

In the case of protons and other ions, the amount of energy deposited in a cell nucleus by a single particle (including all secondary particles) is much higher than in the case of irradiating by photon or electron beams. The number of particles that contribute to a given dose level is lower, and the severity of damages formed by individual particles increases significantly. Even though it is still the collective effect of all the damages that leads to cell inactivation, the role of single-particle effects is much more important than in the case of photon irradiation. Even the impact of a single particle may result in inactivating the cell; compare e.g.\ \citep{Brenner}. Model scheme that enables to represent the effects of single particles as well as their synergetic or saturation combinations is very helpful in such a case.

The basic idea of distinguishing the single-particle and collective effects in the case of proton and light ion irradiation has been formulated by \citet{Judas+Lok-JTBi}, who also suggested that survival curves should be represented by higher-order polynomials instead of their simple parabolic description by the LQ model. Further analyses of the model framework have demonstrated that the polynomial expansion can be used reliably in the low-dose region only. The model has been reformulated in order to ensure its applicability to survival curves measured over a wide range of doses; it has been also developed to greater detail~\citep{PhD}.  Applications of the model to analyses of experimental data have shown the possibility to describe a large variety of experimental cell survival curves. In comparison to phenomenological models, the two-stage model has enabled to represent systematically not only the global shape of cell survival curves but also their detailed structure. The importance of this feature of the model arises from the fact that even small deviations of cell survival curves from their global parabolic or linear shape might be largely amplified in the case of fractionated irradiation.

%%%%%%%%%%%%%%%%%%%%%%%%%%%%%%%%%%%%%%%%%%%%%%%%%%%%%%%%%%%%%%%%%%%%%%%%%%%%%%%%%%%%%

%%%%%%%%%%%%%%%%%%%%%%%%%%%%%%%%%%%%%%%%%%%%%%%%%%%%%%%%%%%%%%%%%%%%%%%%%%%%%%%%%%%%%
\section{Probabilistic two-stage model}
\label{sec:ProbabilisticTwoStageModel}

%%%%%%%%%%%%%%%%%%%%%%%%%%%%%%%%%%%%%%%%%%%%%%%%%%%%%%%%%%%%%%%%%%%%%%%%%%%%%%%%%%%%%
\subsection{Basic model framework}
\label{sec:BasicModelFramework}
Within the probabilistic two-stage model of biological effects of ionizing particles, the following characteristics of the underlying radiobiological mechanism are taken into account: (i) the actual number of primary particles depositing energy to the cell nucleus, (ii) the amount of energy deposited (including, for the sake of simplicity, the energy delivered by all secondary particles), and (iii) the resulting probability of cell inactivation which reflects the complexity of the DNA damage produced.

%%%%%%%%%%%%%%%%%%%%%%%%%%%%%%%%%%%%%%%%%%%%%%%%%%%%%%%%%%%%%%%%%%%%%%%%%%%%%%%%%%%%%
\subsubsection{Number of traversing particles.}
\label{sec:NumberOfTraversingParticles}

The numbers of particles traversing nuclei of individual cells of the irradiated tissue or sample are of stochastic nature. Their average number, $k_{av}$, is given by particle fluence $\Phi$ or applied dose $D$,
\begin{equation}\label{kav}
	k_{av} = hD \ .
\end{equation}
Here, $h$ denotes the proportionality constant, which reflects also geometrical characteristics of the given cell line and irradiation arrangement. The distribution of traversing particle numbers over the cell population can be described then by Poisson statistics,
\begin{equation} \label{WDS-2}
	P_k (D) = \frac{(hD)^k}{k!} \exp(-hD) \ .
\end{equation}
Here we have assumed that the impacts of individual particles can be considered as fully random, and that the diameters of the most efficient track parts are by far smaller than the size of cell nucleus, which is in agreement with the results of track structure studies \citep{Holley}.

%%%%%%%%%%%%%%%%%%%%%%%%%%%%%%%%%%%%%%%%%%%%%%%%%%%%%%%%%%%%%%%%%%%%%%%%%%%%%%%%%%%%%
\subsubsection{Energy transferred to cell nucleus.}
\label{sec:EnergyTransferredToCellNucleus}

The amount of energy transferred to cell nucleus (or, more specifically, to chromosomal system and sensitive regions within it) together with the spatial distribution of individual energy transfer events play key roles with respect to the complexity of the resulting DNA damage. This is in turn crucial for the future fate of the given cell. The amount of energy transferred to chromosomal system by each primary particle (including all its secondaries) is a stochastic quantity. It is influenced by energy and energy-loss spectra of beam particles. It may be, in principle, estimated on the basis of theoretical predictions, Monte Carlo calculations or microdosimetry measurements. Let us denote by $\pi_1(\varepsilon)$ the spectrum of energy $\varepsilon$ deposited to cell chromosomal system in one event (one traversal, i.e.\ one primary particle plus all its secondaries). The normalization condition reads 
\begin{equation} \label{WDS-XX}
	\int_0^{\varepsilon_{max}} \mathrm{d} \varepsilon \ \pi_1(\varepsilon)  =  1 \ ,
\end{equation}
where $\varepsilon_{max}$ stands for the maximum energy that may be transferred in one event. The spectrum of energy transferred in $k$ events is given, then, by a corresponding convolution
\begin{equation} \label{WDS-3}
	\pi_k(\varepsilon) = \Big[ \underbrace{\pi_1 * \pi_1 * \dots * \pi_1}_{k} \Big] (\varepsilon) 
	  = \int_0^{\varepsilon_{max}} \mathrm{d} \varepsilon' \ \pi_1(\varepsilon')
	    \ \pi_{k-1}(\varepsilon-\varepsilon') \ ,
\end{equation}
which fulfils the condition $\int_0^{k \; \varepsilon_{max}} \mathrm{d} \varepsilon \  \pi_k(\varepsilon) = 1$. The average energy deposited per cell nucleus (or chromosomal system), which is obtained by weighting correspondingly the average energy deposited by a given number of particles,
\begin{equation} \label{WDS-YY}
	\varepsilon_{av} = 
	  \sum_k P_k \int_0^{k \; \varepsilon_{max}} \mathrm{d}\varepsilon \ 
	    \varepsilon \ \pi_k(\varepsilon) \ ,
\end{equation}
is proportional to the applied dose $D$ (or to particle fluence $\Phi$).

%%%%%%%%%%%%%%%%%%%%%%%%%%%%%%%%%%%%%%%%%%%%%%%%%%%%%%%%%%%%
\subsubsection{Cell survival probability.}
\label{sec:CellSurvivalProbability}

The number of events (traversals of primary particles), $k$, and the total amount of transferred energy, $\varepsilon$, stand for the primary characteristics describing processes of the physical phase of the radiobiological mechanism. The subsequent chemical and biological processes include large cascades of chemical reactions that lead to the formation of DNA damage, DNA repair processes, and further processes of the biological response of the cell that result finally either in its survival or inactivation. As far as cell survival is concerned, these complex processes can be taken into account in a simplified way by representing the final endpoints directly, i.e.\ by considering only the final probabilities of cell inactivation. In the following we shall denote by $p_k^{(i)}(\varepsilon)$ the average cell inactivation probability after the impact of $k$ (primary) particles that have deposited energy $\varepsilon$. Note that a similar approach, based e.g.\ on DSB-induction probabilities, might be used if other effects, e.g.\ the yields of DNA damage, were the endpoints of interest. Probability of cell survival (i.e., cell survival curve) is, then, given by 
\begin{equation}\label{WDS-4}
	s(D) = 1- \sum_k P_k (D) \int_0^{k \; \varepsilon_{max}} \mathrm{d}\varepsilon \
	                      \pi_k(\varepsilon) \ p_k^{(i)}(\varepsilon)  \ ,
\end{equation}
where the distribution of the actual number of particle traversals, $k$, and also the spectra of transferred energy, $\pi_k(\varepsilon)$, have been taken into account. Here the dose dependence is included in $P_k(D)$, see Eq.~(\ref{WDS-2}). Note that the average number of traversing particles per unit dose, $h$, as well as transferred energy spectra, $\pi_k(\varepsilon)$, and also inactivation probabilities $p_k^{(i)}(\varepsilon)$ vary significantly for different particles, their LET values and/or diverse cell lines. This is in agreement with the experimental evidence of the dependence of cell survival on all these factors. Given the inactivation probabilities $p_k^{(i)}(\varepsilon)$ together with the distribution of particle numbers $P_k$ and the spectra of transferred energy $\pi_k(\varepsilon)$ under the given experimental or clinical irradiation conditions, one would be able to calculate the cell survival probability according to Eq.~(\ref{WDS-4}). Using corresponding model approaches \citep{Webb+Nahum, Niemierko+Urie+Goitein, Niemierko+Goitein}, macroscopic endpoints such as the tumor control and normal tissue complication probabilities (TCP and NTCP) can be estimated, then. In evaluating the cell survival probability according to Eq.~(\ref{WDS-4}), a crucial role is played by the inactivation probabilities $p_k^{(i)}$. These probabilities reflect the biological effectiveness of different radiation modalities. They may be interpreted directly in terms of physical and biological processes. They represent, therefore, a solid basis for establishing more detailed microscopic models of radiobiological mechanism. Methods of evaluating these probabilities on the basis of analyzing corresponding experimental cell-survival data will be derived and discussed in the next sections.

%%%%%%%%%%%%%%%%%%%%%%%%%%%%%%%%%%%%%%%%%%%%%%%%%%%%%%%%%%%%%%%%%
\subsection{Monoenergetic ions}
\label{sec:MonoenergeticIons}
Radiobiological experiments aimed at establishing the biological effectiveness of different radiation kinds are performed usually by irradiating cell monolayers by monoenergetic beams only. The beam energy is chosen so as to correspond to certain regions in the Bragg peak of the given particles (so-called track-segment experiments). In such a case the spectrum of transferred energy is rather narrow, at least as compared to non-monoenergetic beams. The (average) energy transferred to cell nuclei is proportional to the LET value $\lambda$ of the particles. Moreover, in this case the average number of (primary) particles traversing cell nuclei, i.e.\ the parameter $h$ (cf.~Eqs.~(\ref{kav}) and~(\ref{WDS-2})), can be related to the effective cross-section, $\sigma$, of the cell nucleus (or chromosomal system) and LET value $\lambda$ by
\begin{equation}\label{WDS-5}
	h=\frac{C\sigma}{\lambda} \ .
\end{equation}
Approximating the density of traversed medium by that of water, the conversion constant $C = 6.24 \ \mathrm{keV \; Gy^{-1} \; \mu m^{-3}}$ for the usual choice of units ($\lambda \mathrm{[keV/\mu m]}$, $D\mathrm{[Gy]}$, $\sigma \mathrm{[\mu m^2]}$). Note that $\sigma$ denotes the geometrical effective cross-section of the chromosomal system (or of sensitive region within cell nucleus) to be traversed by the given particles. The consequent processes of DNA damage and cellular response are not involved in $\sigma$, but treated separately. Note especially that $\sigma$ is not identical to cell inactivation cross-section, which is used by some authors to describe (the linear component of) the probability, per 1 particle, of the cell to be inactivated; compare e.g.~\citep{LEM-1}.

Cell survival probability can be, then, expressed as
\begin{equation}\label{WDS-6}
	s(D) = \sum_k P_k(D,\lambda) \left( 1-p_k^{(i)}(\lambda) \right) 
	     = 1 - \sum_k P_k(D,\lambda) \ p_k^{(i)}(\lambda)  \ ,
\end{equation}
where (average) inactivation probabilities $p_k^{(i)}(\lambda)$ are now expressed as functions of LET value $\lambda$ instead of energy $\varepsilon$; they are given by $p_k^{(i)}(\lambda)=\int_0^{k\varepsilon_{max}} \mathrm{d}\varepsilon \ \pi_k(\varepsilon) \ p_k^{(i)}(\varepsilon)$.

%%%%%%%%%%%%%%%%%%%%%%%%%%%%%%%%%%%%%%%%%%%%%%%%%%%%%%%%%%%%%
\subsection{Detailed model of inactivation probabilities}
\label{sec:DetailedModelOfInactivationProbabilities}

In the case of particles characterized by higher LET values, i.e.\ protons, neutrons and ions, even the impact of a single primary particle (including the effects of all its secondary particles) may produce a severe damage to chromosomal DNA, which might finally lead to cell inactivation. In the following, we shall refer to these damages as to "single-particle induced" damages. Let us denote the probability of their induction, per one traversing particle, by $a$. If $k$ particles have traversed the chromosomal system, the probability of not being damaged in this way is given by
\begin{equation}
	\widetilde{q}_k^A = (1-a)^k \ .
\end{equation}

A vast majority of DNA damages formed by traversing particles are not of this severity; they can be repaired successfully and do not lead to cell inactivation. However, DNA damages caused by different particles might combine to form a kind of damage that is difficult to be repaired. Synergetic combinations of such damages might include combinations of single-strand breaks (SSBs) into a double-strand break (DSB) or DSBs into a more complex damage. They need not be limited to one chromosome only; e.g.\ the impact of a particle pair damaging corresponding segments of chromosomes of a given homologous pair. To account for these combined effects, let us denote by $b$ the average probability that such a severe damage, which alone does not result in inactivating the cell, has been formed by a single particle. If this is the only damage to the cell, we shall assume that it can be repaired quite easily and the cell will survive, possibly with some delay in its proliferation cycle. However, combination of two or more such damages might lead to cell inactivation if not repaired correctly. Probability that no such "pair" or "combined" damage\footnote{Note that also the probability of inducing "pair damage" to DNA refers to the effect of a single particle. The term "pair" or "combined" damage has been chosen just to highlight the lower severity of such damage, i.e.\ the necessity of being combined with another damage of a similar character in order to lead to cell inactivation.} has been formed after the impact of $k$ particles is then, to the first approximation, given by
\begin{equation}
	\widetilde{q}_k^B = (1-b^2)^{k(k-1)/2}   \ .
\end{equation}
Here we have assumed, for the sake of simplicity, that the synergetic effects occur predominantly from particle pairs only; the exponent in the last formula stands just for the number of particle pairs. 

The final biological effect, measured e.g.\ by means of cell survival, does not depend only on the initial yield of DNA damages. Cellular repair processes have been shown to contribute significantly to this point, too. In the following we shall assume that severe DNA damages potentially leading to cell inactivation, either caused by single particles ("single-particle induced" damages) or combined from multi-particle effects ("combined" or "pair" damages), may be repaired by the cell. We shall denote the probabilities of repairing them successfully by $r^A$ and $r^B$. After the cellular repair processes took place, the probability that a residual severe damage is present equals, then,
\begin{equation}
	p_k^A  =  ( 1-\widetilde{q}_k^A ) ( 1 - r_k^A )  =  \left( 1 - (1-a)^k \right)  \left( 1 - r_k^A \right)
\end{equation}
in the case of single-particle induced damages, and
\begin{equation}
	p_k^B  =  ( 1-\widetilde{q}_k^B ) ( 1 - r_k^B )  =  \left( 1 - (1-b^2)^{k(k-1)/2} \right)  \left( 1 - r_k^B \right) 
\end{equation}
in the case of combined (or "pair") damages, respectively. Here we have explicitly indicated that the rate and fidelity of repair processes in the two cases need not be equal. The damage complexity increases with the number of traversing particles $k$ and with the LET value $\lambda$. Therefore, the probability of successful repair of DNA damages should decrease with these factors. The repair probability might, however, in some cases exhibit more complex behavior, corresponding to the onset of cellular repair processes.

The probabilities that the mentioned damages have either not been formed at all or have been repaired faithfully are just the complements of the last terms,
\begin{equation}
	q_k^A = 1-p_k^A  =  1  -  \left( 1 - (1-a)^k \right)  \left( 1 - r_k^A \right)
\end{equation}
and
\begin{equation}
	q_k^B = 1-p_k^B  =  1  -  \left( 1 - (1-b^2)^{k(k-1)/2} \right)  \left( 1 - r_k^B \right)  \ ,
\end{equation}
respectively. If the cell should survive, it must not be inactivated by either of the mentioned mechanisms; i.e., neither single-particle induced nor pair damages must be present after the repair processes took place. The cell survival probability is, therefore, given by
\begin{equation}
	q_k \, = \, q_k^A \ q_k^B \ ,
\end{equation}
and the inactivation probability
\begin{equation}
	p_k^{(i)} = 1 - q_k \ .
\end{equation}

The cell survival probability $s(D)$ after applied dose $D$ is given then by Eq.~(\ref{WDS-6}), where the distribution of the numbers of particles traversing individual cell nuclei is given by Poisson distribution, Eq.~(\ref{WDS-2}). In the case of Bragg peak or other non-monoenergetic irradiation, the spectra of transferred energy have to be taken into account, too, according to Eq.~(\ref{WDS-4}). Similarly, in the case of mixed beam irradiation one has to convolve these results with the spectrum of traversing particles (i.e.\ with the abundances of individual ion species in the beam), again in the way analogous to that indicated in Eq.~(\ref{WDS-4}).

The probabilities of "single-particle" and "pair" damage formation, $a$ and $b$, as well as of successful repair, $r_k^A$ and $r_k^B$, depend significantly on the radiation quality (ion kind and LET value) and, of course, on the biological characteristics of given cell line (cellular radiosensitivity, cell cycle phase etc.). The values of these parameters may be established by analyzing corresponding experimental data. Systematic sets of data, in which cell survival has been assessed for a given ion at different energies (different LET values), might be used to evaluate even the functional dependence of these probabilities on particles' LET values. Corresponding analyses have been performed by us for various ions; the results will be shown in the following sections. In determining the values of the damage induction and repair probabilities, more detailed microscopic models, representing the underlying chemical and biological processes, might be helpful, too. In fact, the given model scheme represents a framework enabling one to relate the outputs of those microscopic models to cell survival characteristics, which might be assessed experimentally in a direct way.

The probability of successful repair, $r$, should reflect the complexity of the total damage formed by all the traversing particles. The damage complexity might be, to the first approximation, estimated from the total amount of energy deposited to cell nucleus (or chromosomal system). For the sake of simplicity, we have therefore limited ourselves to consider the repair probability, $r(k,\lambda)$, to be a function of the transferred energy only, $\varepsilon \sim k.\lambda$:
\[ r(k,\lambda) \rightarrow r(k.\lambda) \ . \]

%%%%%%%%%%%%%%%%%%%%%%%%%%%%%%%%%%%%%%%%%%%%%%%%%%%%%%%%%%%%%
\subsection{Basic model features}
\label{sec:BasicModelFeatures}

\begin{figure}[!htb]
	%\centering
		\includegraphics[trim=-0.3cm -0.3cm -0.3cm -0.3cm, clip, angle=-90, width=0.40\textwidth]
{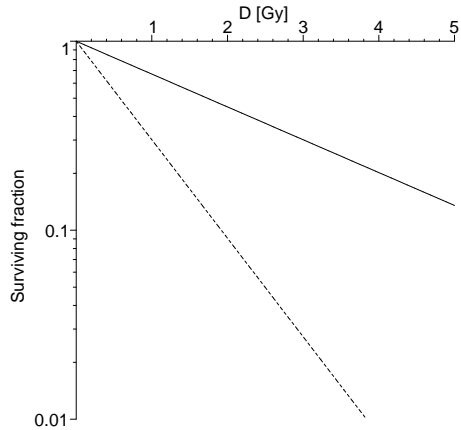}
	\caption{Examples of linear cell survival curves, obtained by the probabilistic two-stage model by putting $h=2$; $a=0.2$ (solid line) or 0.5 (dashed line); $b = 0$; $r^A = 0 = r^B$. Note that the scaling of the cell survival with applied dose is somehow arbitrary in these examples; it corresponds to the choice of $h$ value.}
	\label{fig:linear}
\end{figure}

\begin{figure}[!htb]
	%\centering
		\includegraphics[trim=-0.3cm -0.3cm -0.3cm -0.3cm, clip, angle=-90, width=0.40\textwidth]
{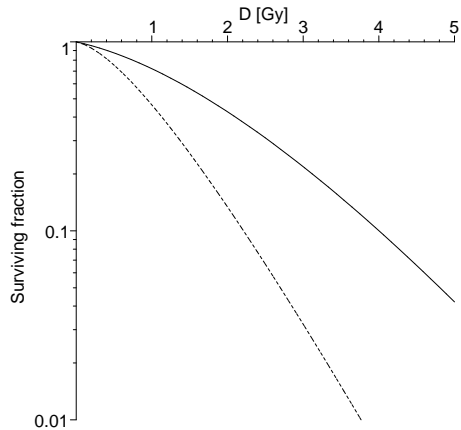}
	\caption{Parabolic cell survival curves corresponding to $h = 2$; $a = 0.1$; $b = 0.3$ (solid line) or 0.8 (dashed line); repair not taken into account, $r^A = 0 = r^B$.}
	\label{fig:parabolic}
\end{figure}

\begin{figure}[!htb]
	\centering
	\begin{minipage}[t]{0.40\textwidth}	
		\includegraphics[angle=-90, width=1.00\textwidth]{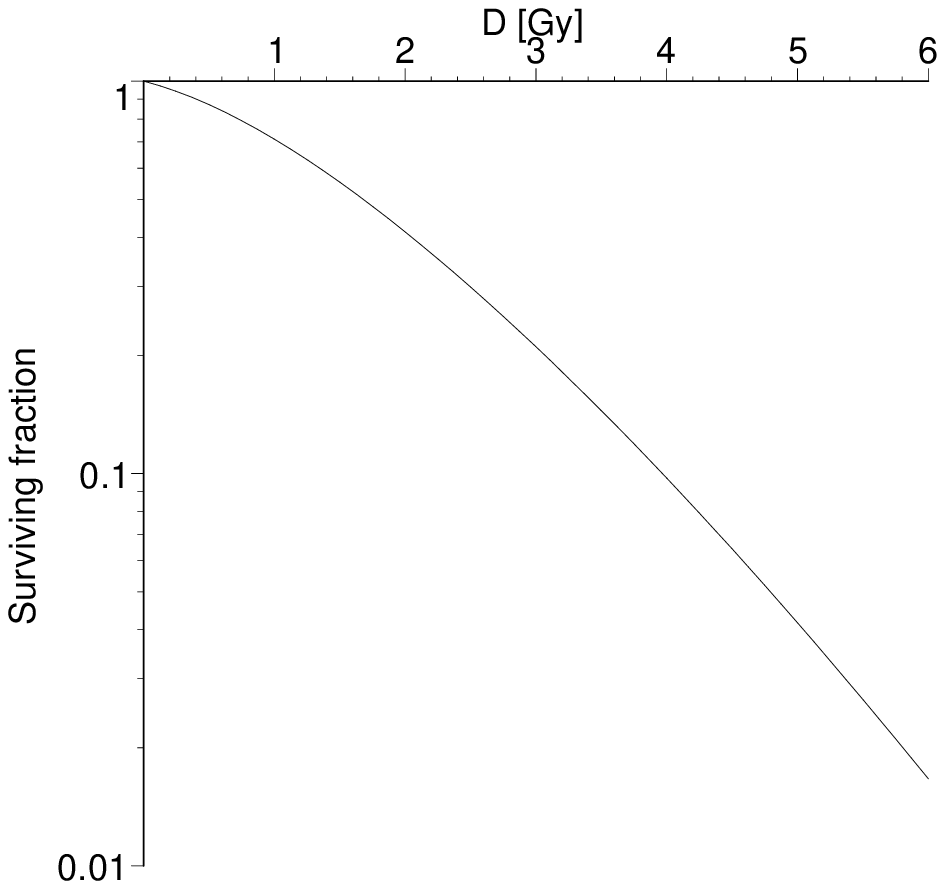}
	\end{minipage}
	\begin{minipage}[t]{0.05\textwidth}
		\rule[-4pt]{0pt}{16pt}
	\end{minipage}
	\begin{minipage}[b]{0.40\textwidth}
		\includegraphics[angle=-90, width=1.00\textwidth]{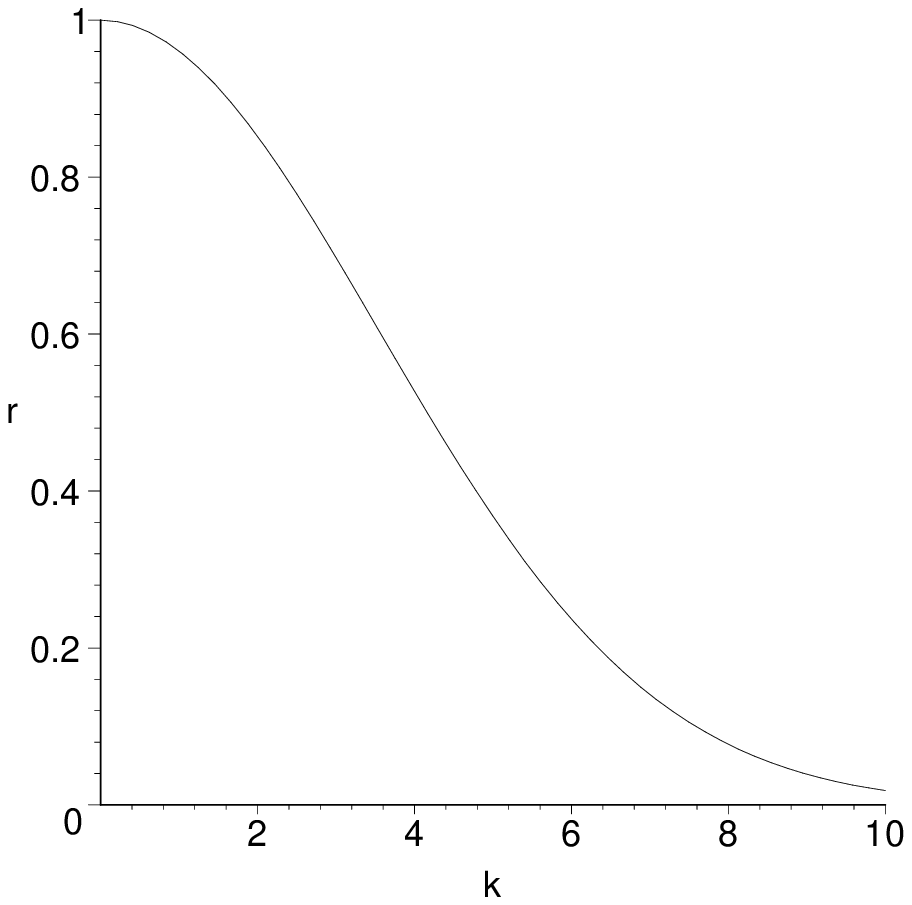}
	\end{minipage}
	\caption{An example of survival curve characterized by a global parabolic shape even if repair taken into account. Damage induction probabilities $a = 0.1$; $b = 0.8$; number of particles traversing the nucleus per unit dose $h = 2$. Repair of combined damages as shown in the right-hand panel; parameterized by auxiliary function $r(k)=\exp(-(r_1k)^{r_2})$, where $r_1=0.2$, $r_2=2$. Repair of single-particle induced damages not taken into account, $r^A = 0$.}
	\label{fig:parabolic+repair}
\end{figure}

Let us discuss the basic features of the given model framework, especially the role of individual model parameters in determining the shapes of cell survival curves. Obviously, there are no difficulties in describing simple linear or parabolic cell survival curves by the given model scheme. Linear curves (Fig.~\ref{fig:linear}) can be represented by taking into account the terms including $p_k^A$ only, i.e.\ by assuming that all the lesions formed by the traversing particles are lethal, do not need to be combined with damages caused by other particles (i.e.\ $b = 0$), and cannot be repaired by the cell ($r = 0$). This is in agreement with experimental evidence: Linear survival curves have been found especially in very sensitive cell lines and in cell lines that have been proved to be deficient in repair~\citep{Steel-Joiner}. Linear response has been observed also in irradiating by very high-LET particles, where the severity of DNA damages is assumed to be very high~\citep{Kraft-WhatKindOfRBShouldBeDone}. Similarly, survival curves characterized by a global parabolic shape can be represented by taking into account the terms including $p_k^B$; compare Fig.~\ref{fig:parabolic}. Even cell lines exhibiting DNA repair of a certain level may be characterized by roughly parabolic survival curves; compare Fig.~\ref{fig:parabolic+repair}.

\begin{figure}[!htb]
	\centering
	\begin{minipage}[t]{0.40\textwidth}	
		\includegraphics[angle=-90, width=1.00\textwidth]{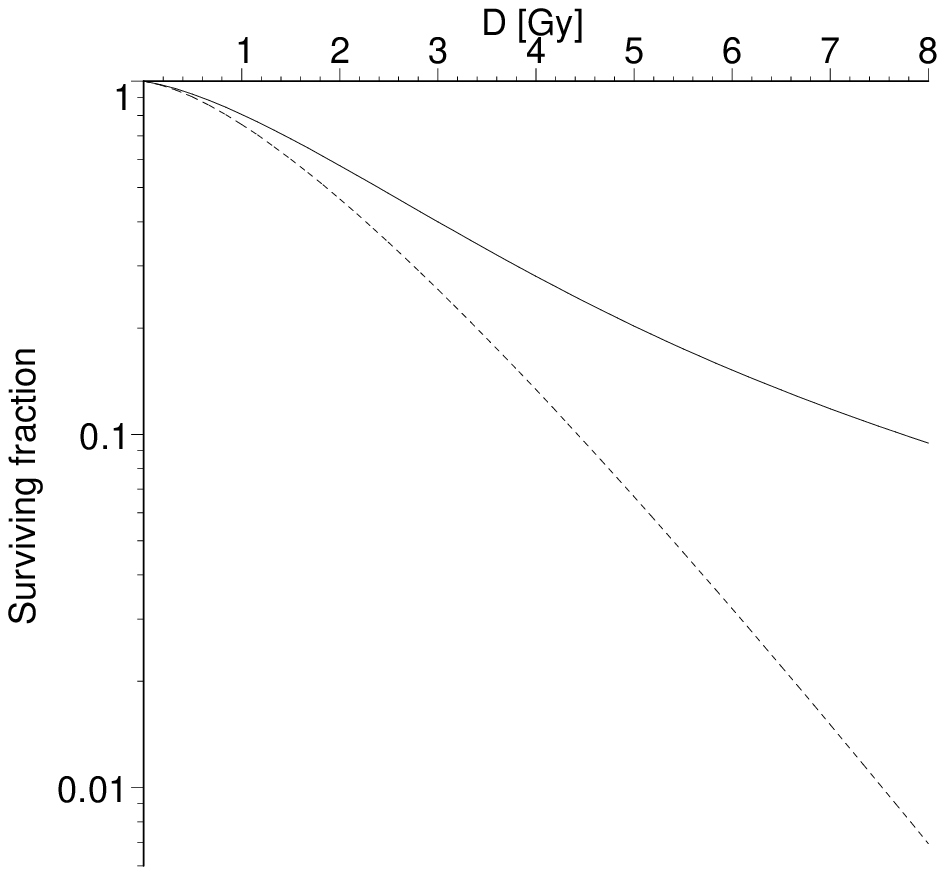}
	\end{minipage}
	\begin{minipage}[t]{0.05\textwidth}
		\rule[-4pt]{0pt}{16pt}
	\end{minipage}
	\begin{minipage}[b]{0.40\textwidth}
		\includegraphics[angle=-90, width=1.00\textwidth]{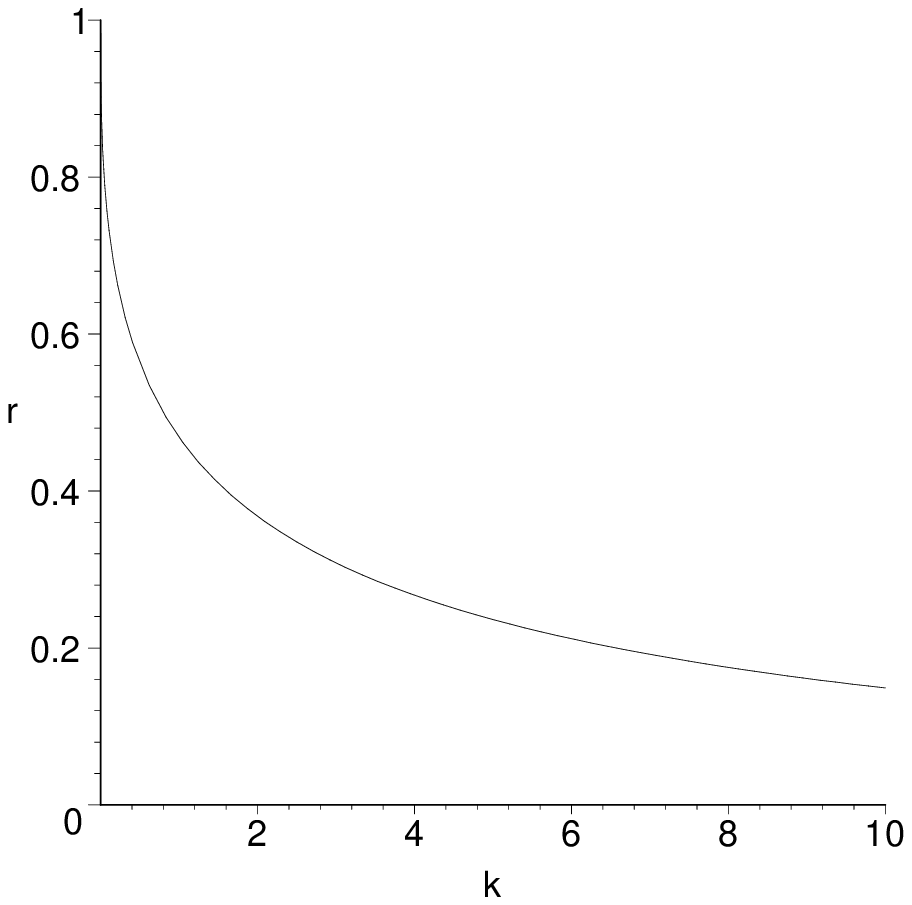}
	\end{minipage}
	\caption{Example of more complex behavior of cell survival curves. Solid line obtained by putting $h=1$, $a=0.1$, $b=0.8$. Repair probability, decreasing with the number of particles traversing the nucleus, as depicted in the right-hand panel; parameterized by auxiliary function $r(k)=\exp(-(r_1k)^{r_2})$, where $r_1=0.5$, $r_2=0.4$. Survival curve shown by dashed line corresponding to the same parameter values but without repair, $r^B=0$; in both the cases being put $r^A=0$.}
	\label{fig:complex_shape}
\end{figure}

Cell survival curves may, however, exhibit more complex behavior. The given model scheme can be used to represent survival curves showing, e.g., a downward bending in the region of lower doses followed by an upward bending at higher doses, compare Fig.~\ref{fig:complex_shape}. The corresponding repair probability, to which such complex behavior has to be attributed, is shown, too.

\begin{figure}[!htb]
	\centering
	\begin{minipage}[t]{0.40\textwidth}	
		\includegraphics[angle=-90, width=1.00\textwidth]{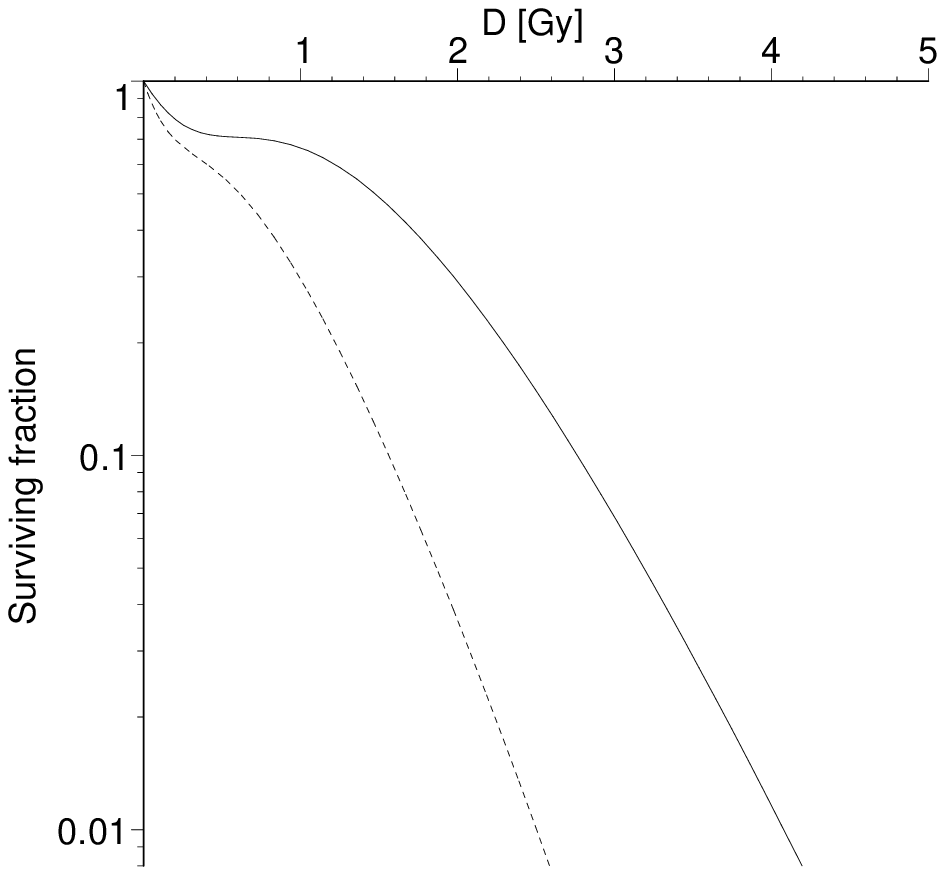}
	\end{minipage}
	\begin{minipage}[t]{0.05\textwidth}
		\rule[-4pt]{0pt}{16pt}
	\end{minipage}
	\begin{minipage}[b]{0.40\textwidth}
		\includegraphics[angle=-90, width=1.00\textwidth]{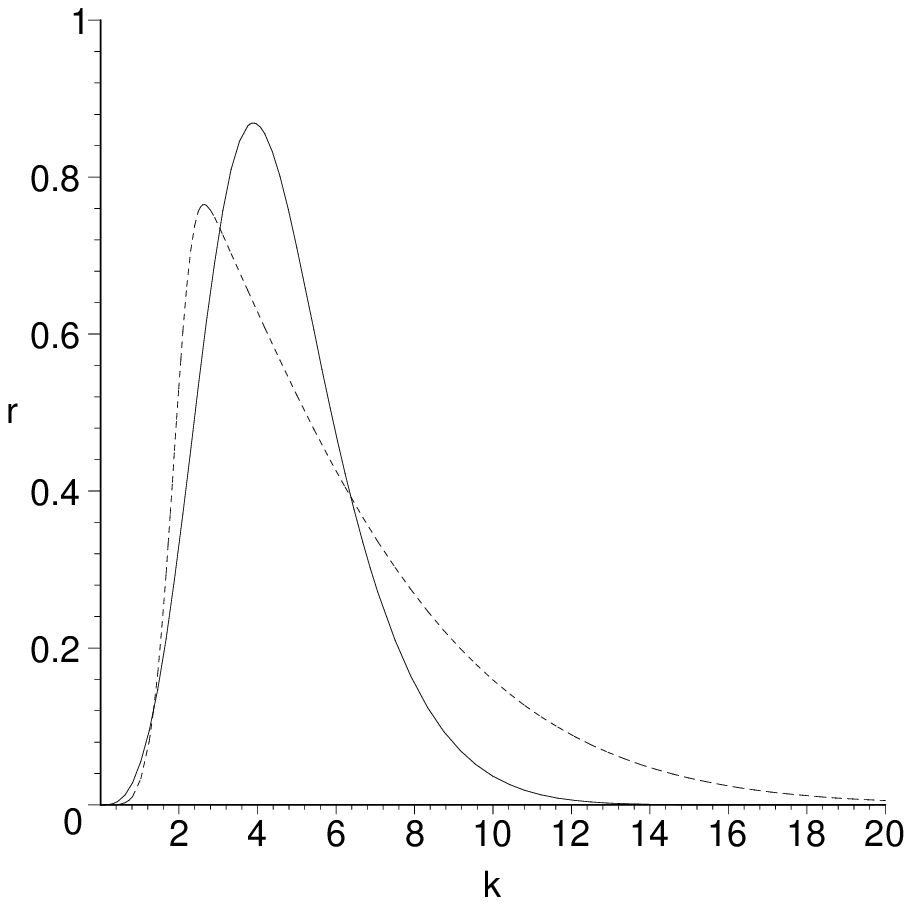}
	\end{minipage}
	\caption{Examples of survival curves exhibiting low-dose hypersensitivity phenomenon. Here $h=4$, $a=0.4$, $b=0$ (solid line) and $h=6$, $a=0.5$, $b=0.2$ (dashed line), respectively. The repair probability $r^A$ (right panel) is characterized by onset behavior, with maximal repair rate in the region of a few traversing particles. Parameterized by auxiliary function $r(k)=\exp(-(r_1k)^{r_2}) [1-\exp(-(r_3k)^{r_4})]$, where $r_0=2$, $r_1=0.2$, $r_2=2$, $r_3=0.3$, $r_4=3$ (solid line) or $r_0=1$, $r_1=0.15$, $r_2=1.5$, $r_3=0.5$, $r_4=5$ (dashed line); $r^B=0$ in both the cases.}
	\label{fig:hypersensitivity}
\end{figure}

\begin{figure}[!htb]
	\centering
	\begin{minipage}[t]{0.40\textwidth}	
		\includegraphics[angle=-90, width=1.00\textwidth]{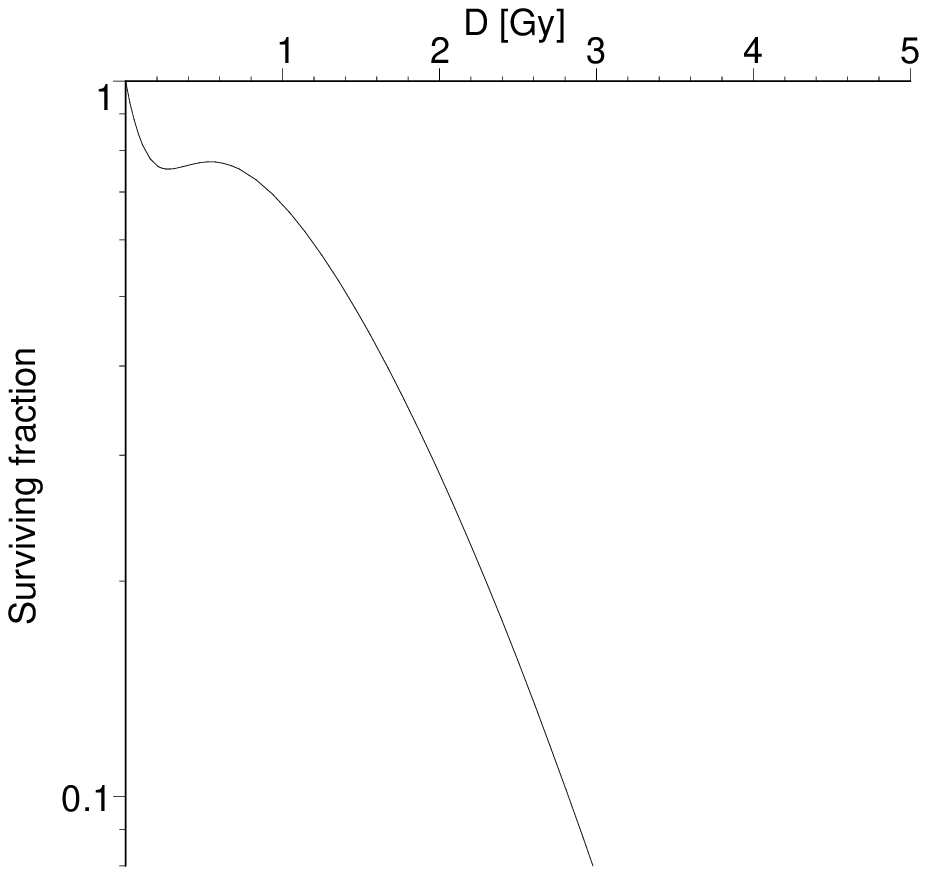}
	\end{minipage}
	\begin{minipage}[t]{0.05\textwidth}
		\rule[-4pt]{0pt}{16pt}
	\end{minipage}
	\begin{minipage}[b]{0.40\textwidth}
		\includegraphics[angle=-90, width=1.00\textwidth]{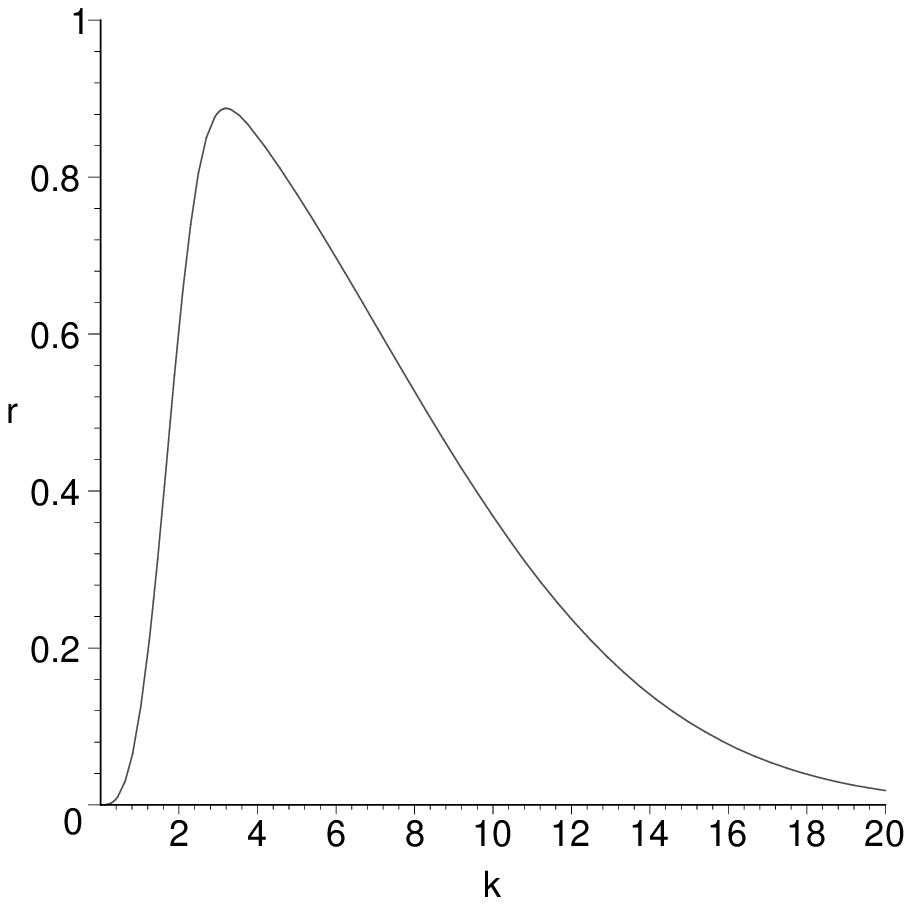}
	\end{minipage}
	\caption{Example of model representation of survival curve exhibiting highly pronounced hypersensitive behavior, with a temporary increase following the low-dose hypersensitivity region. Probability of single-particle induced DNA damage $a = 0.5$, average number of traversing particles per unit dose $h = 6$. Combined damages not taken into account, $b = 0$, $r^B = 0$. Auxiliary parameterization of the repair probability (right panel) $r(k)=\exp(-(r_1k)^{r_2}) [1-\exp(-(r_3k)^{r_4})]$, where $r_0=1$, $r_1=0.1$, $r_2=2$, $r_3=0.5$, $r_4=3$.}
	\label{fig:hypersensitivity-temporary_increase}
\end{figure}

The model enables one to describe also survival curves exhibiting the low-dose hypersensitivity phenomenon, which has been observed for many cell lines irradiated by different particles~\citep{Joiner, Chalmers, Schettino}. In Fig.~\ref{fig:hypersensitivity} two examples of calculated survival curves with hypersensitivity regions are demonstrated. The corresponding repair characteristics, showing an onset (triggering) effect, are displayed, too. Note that if the repair profile exhibits significant variations in the rather narrow range of a few particles traversing the nucleus, the corresponding survival curve may be characterized even by a temporary increase in the medium-dose region (Fig.~\ref{fig:hypersensitivity-temporary_increase}). Such behavior has been found in several experiments, compare e.g.~\citep{Chalmers}.

The examples shown in this section demonstrate clearly that in determining the shape of cell survival curves the repair characteristics may be even more important than the damage induction itself.

%%%%%%%%%%%%%%%%%%%%%%%%%%%%%%%%%%%%%%%%%%%%%%%%%%%%%%%%%%%%%%%%%%%%%%%%
\section{Application to experimental data}
\label{sec:ApplicationToExperimentalData}

The proposed model has been applied to analysis of experimental cell survival data for Chinese hamster V79 cells after irradiation by low-energy protons, which was assessed experimentally by \citet{Belli}. Mono-energetic proton beams were used in their study, with the energies in the range of 0.57~--~5~MeV, corresponding to LET values 7.7~--~37.8~$\mathrm{keV/\mu m}$.

The following parameterization has been used for representing the damage induction and repair probabilities:
\begin{eqnarray}\label{eq:proton}
	a(\lambda) &=& \Big(a_1 \lambda + a_2 \lambda^2\Big) \Big[1-\exp\Big(-(a_3\lambda)^{a_4}\Big)\Big]  \ , \nonumber\\
	b(\lambda) &=& \frac{1 - \exp(-(b_1 \lambda)^{b_2})} {1 + b_3 \exp(-(b_4\lambda)^{b_5})}  \ , \\
	r^B(k,\lambda) &=& \frac{1-\exp(-(r_1 k\lambda)^{r_2})} {1+r_3 \exp(-(r_4 k\lambda)^{r_5})} \ . \nonumber
\end{eqnarray}
Flexible monotonous test functions were used in $b(\lambda)$ and $r^B(k,\lambda)$, whereas $a(\lambda)$ has been allowed to possess a non-monotonous behavior.

\begin{figure}[!htb]
	\centering
	\begin{minipage}[t]{0.4\textwidth}
		\includegraphics[trim=0cm 0.5cm 0cm 1.4cm, clip, angle=-90, width=1.00\textwidth]
		{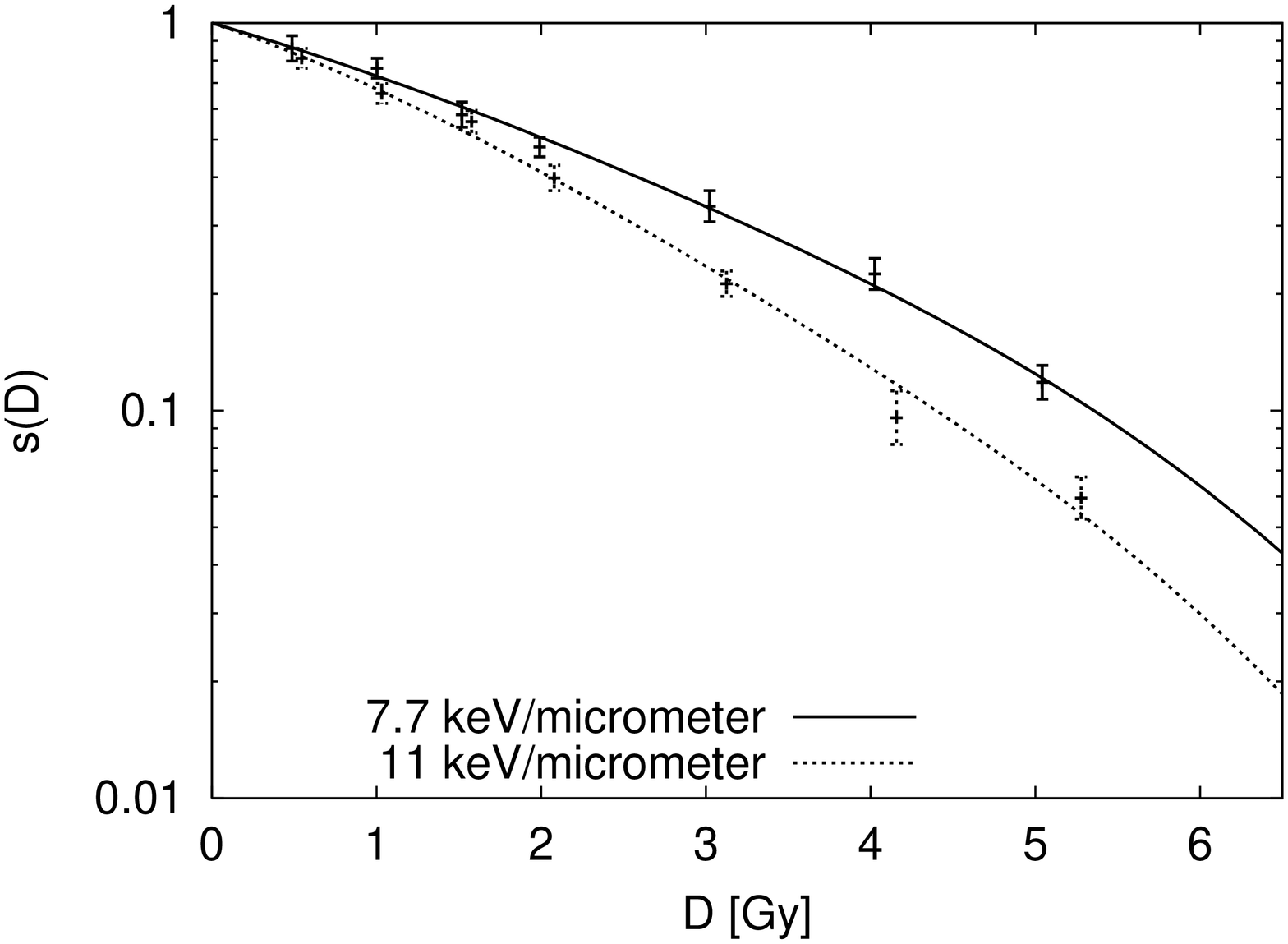}
		\includegraphics[trim=0cm 0.5cm 0cm 1.4cm, clip, angle=-90, width=1.00\textwidth]
		{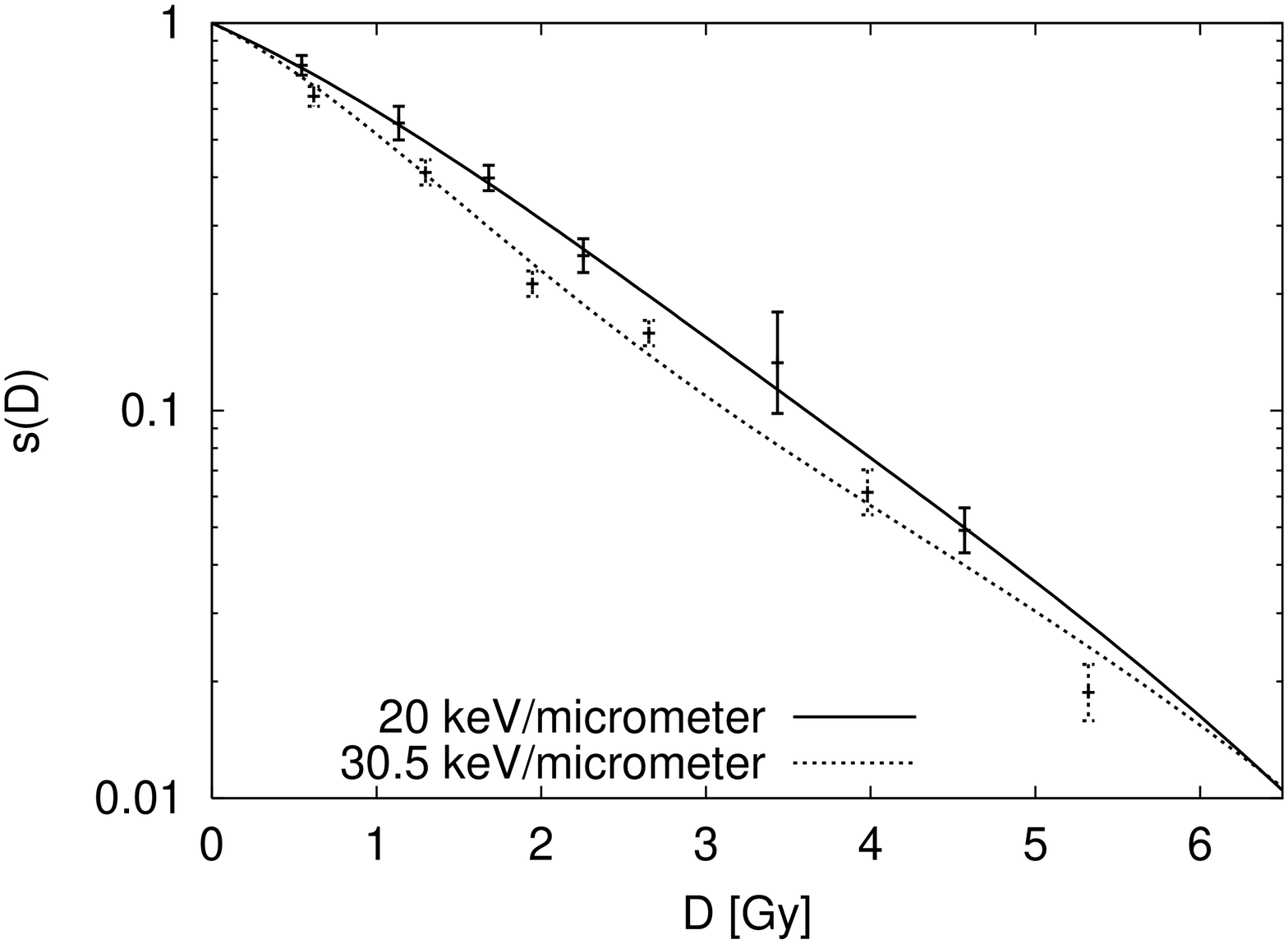}
		\includegraphics[trim=0cm 0.5cm 0cm 1.4cm, clip, angle=-90, width=1.00\textwidth]
		{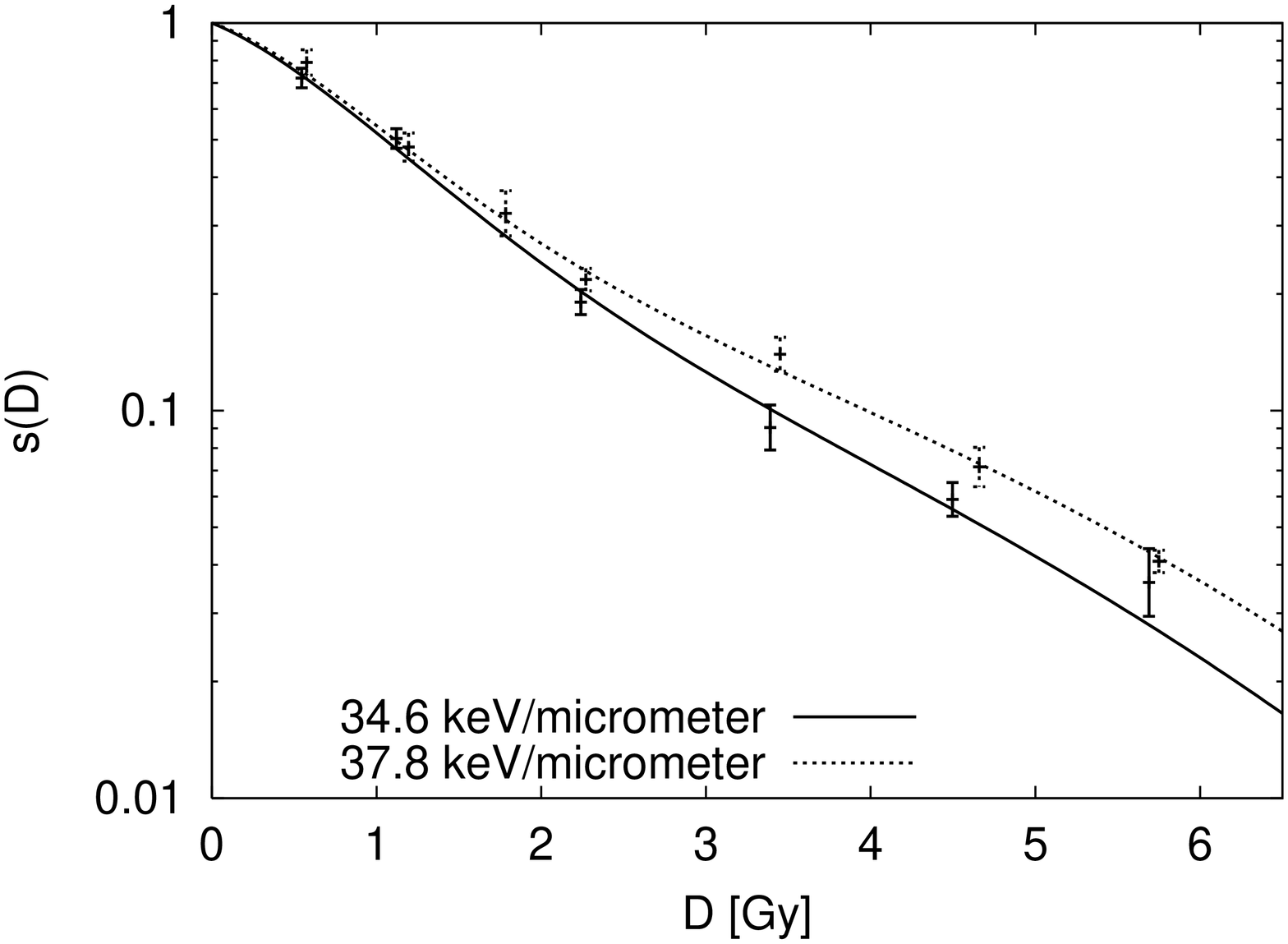}
	\end{minipage}
%	\begin{minipage}[t]{0.01\textwidth}
%		\rule[-4pt]{0pt}{16pt}
%	\end{minipage}
	\begin{minipage}[t]{0.4\textwidth}
		\includegraphics[trim=0cm 0.5cm 0cm 1.4cm, clip, angle=-90, width=1.00\textwidth]
		{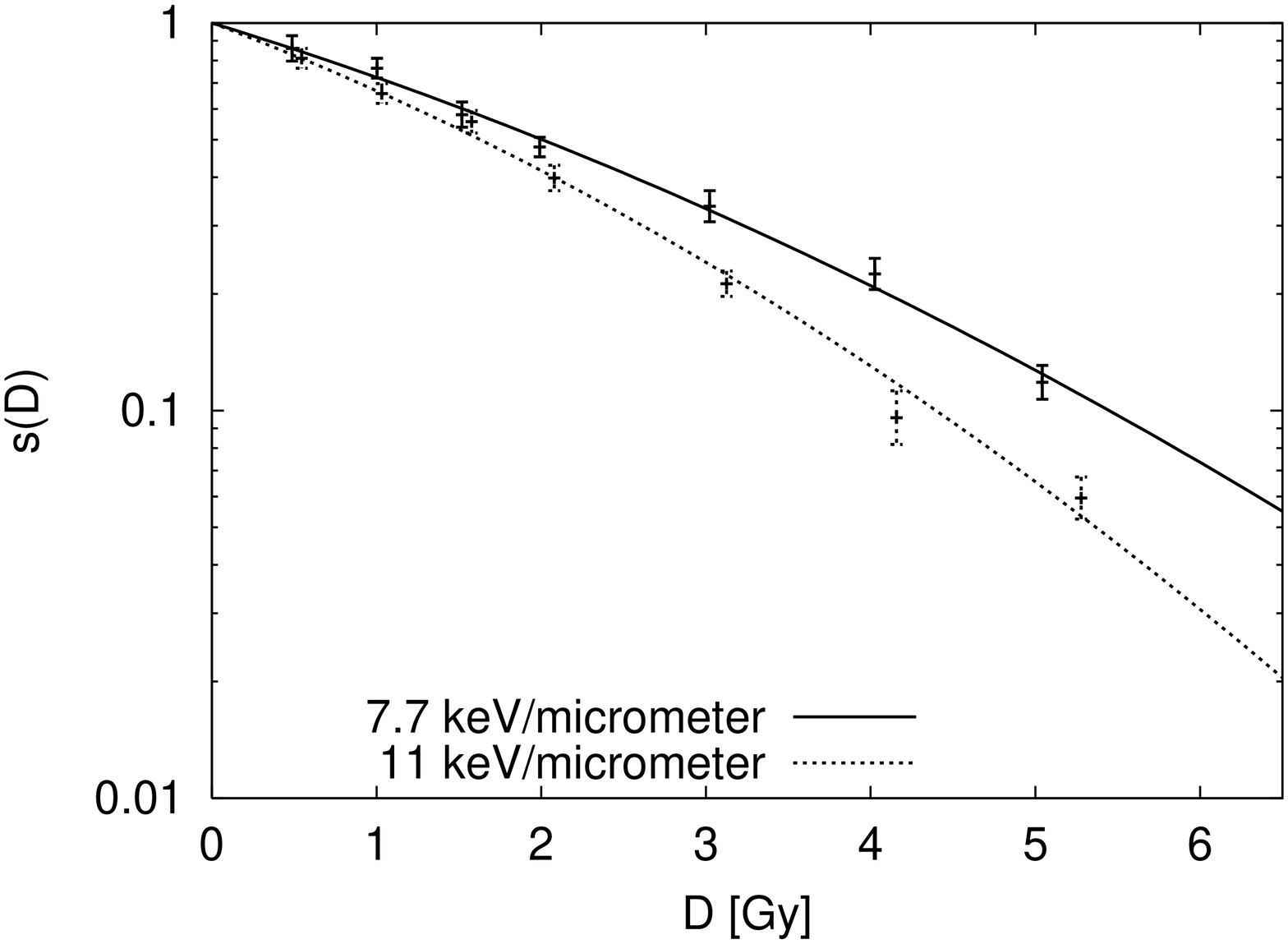}
		\includegraphics[trim=0cm 0.5cm 0cm 1.4cm, clip, angle=-90, width=1.00\textwidth]
		{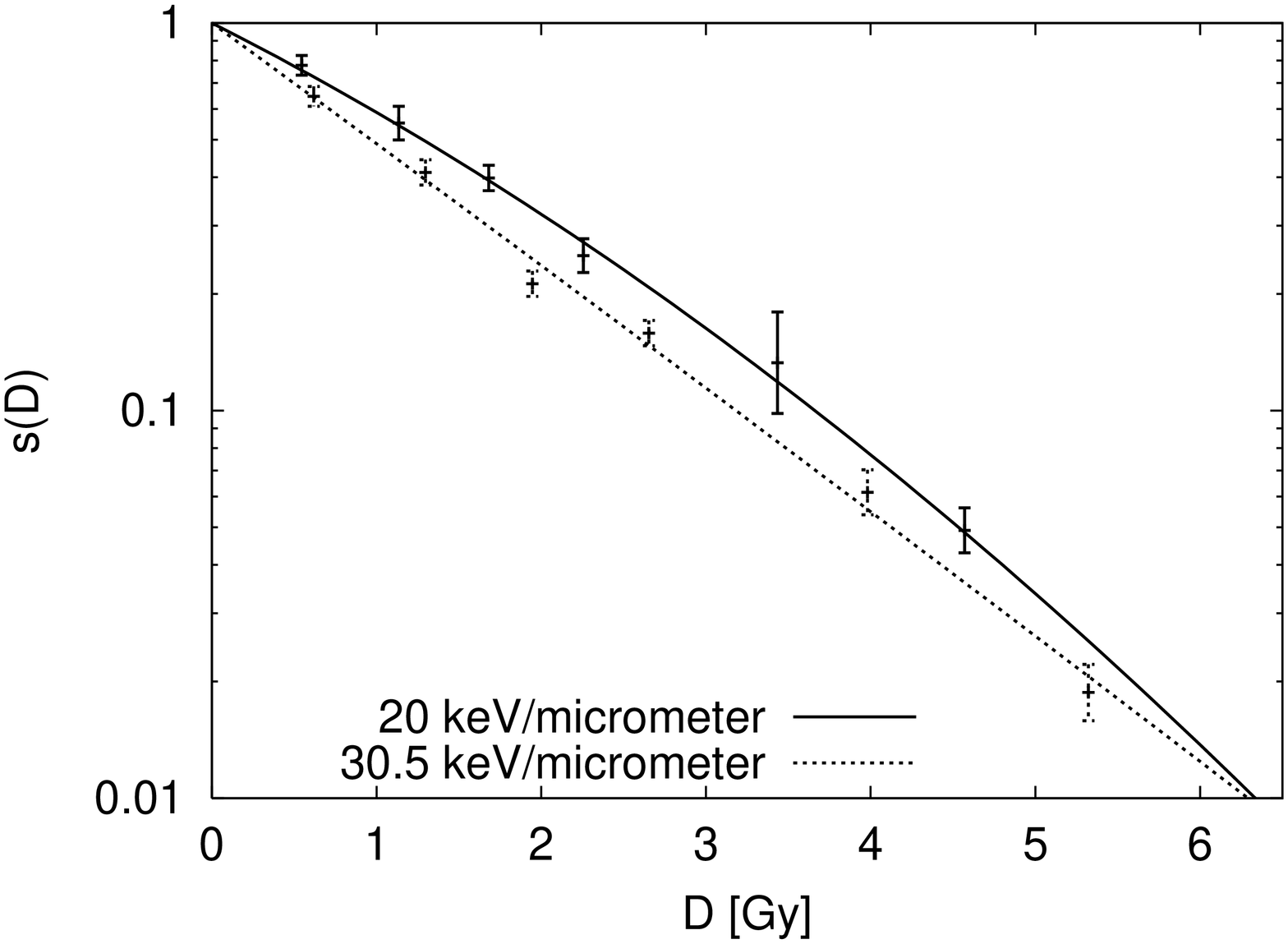}
		\includegraphics[trim=0cm 0.5cm 0cm 1.4cm, clip, angle=-90, width=1.00\textwidth]
		{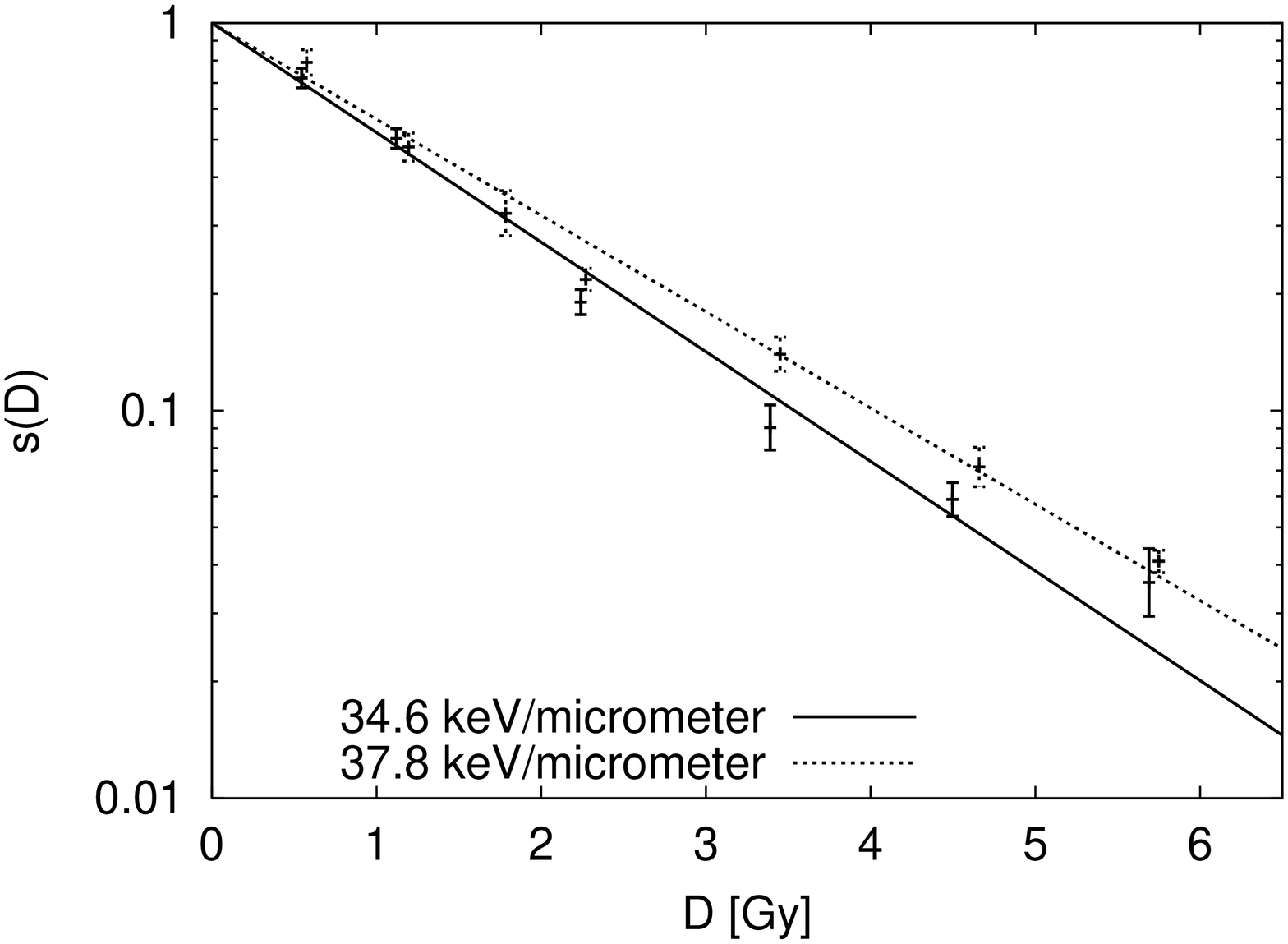}
	\end{minipage}
	\caption{Survival curves for V79 cells irradiated by low-energy protons. Experimental data established by~\citet{Belli}, interpreted on the basis of a systematic analysis using the detailed inactivation scheme of the probabilistic two-stage model (left panel). The corresponding probabilities of DNA damage formation and repair are shown in Figure~\ref{fig:proton-abr}. Fits by the LQ model shown, too (right panel).}
	\label{fig:proton}
\end{figure}

\begin{figure}[!htb]
	\centering
	\begin{minipage}[t]{0.4\textwidth}	
		\includegraphics[angle=-90, width=1.00\textwidth]{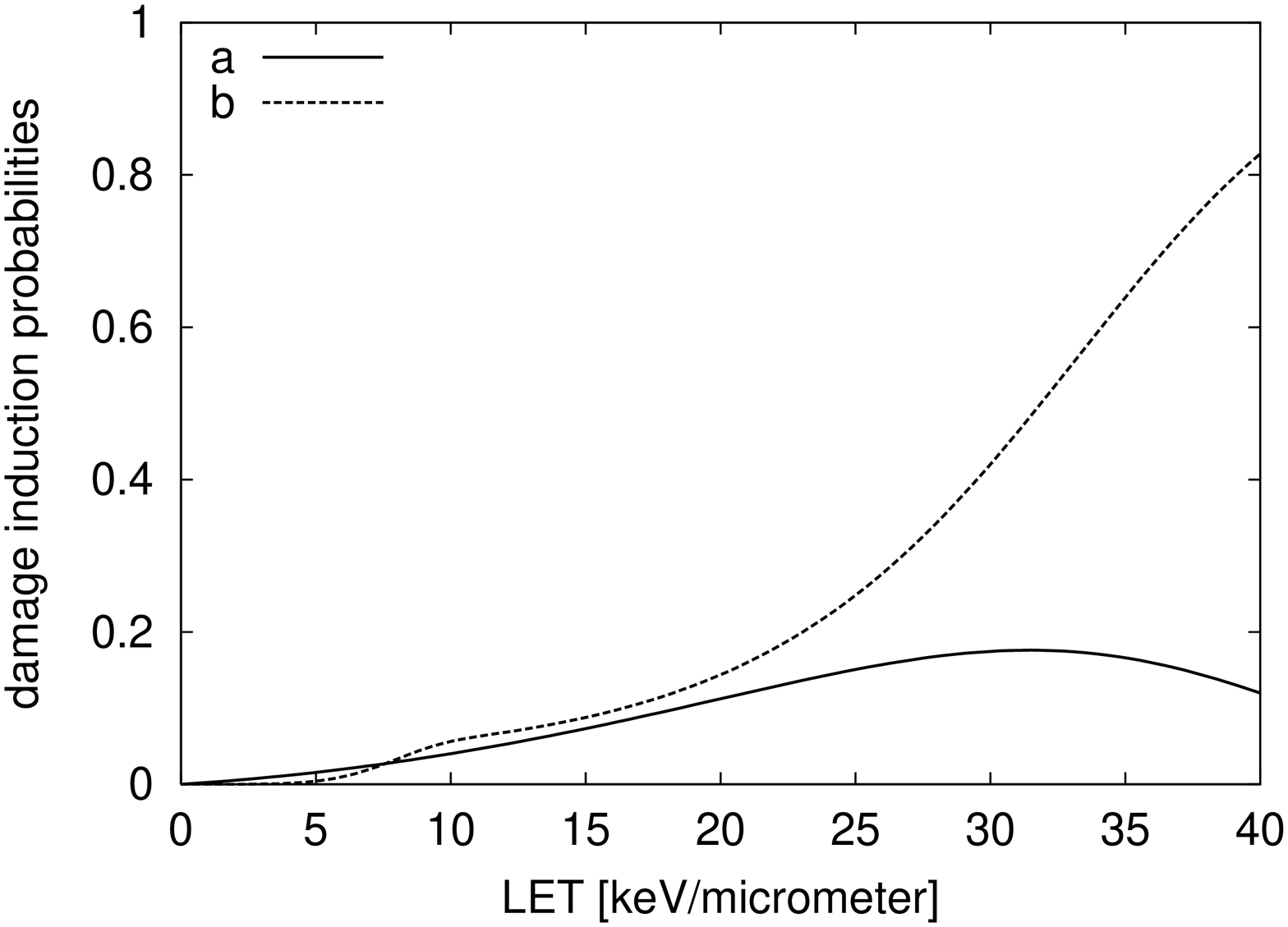}
	\end{minipage}
	\begin{minipage}[t]{0.05\textwidth}
		\rule[-4pt]{0pt}{16pt}
	\end{minipage}
	\begin{minipage}[b]{0.4\textwidth}
		\includegraphics[trim=1cm 2cm 1.5cm 2cm, clip, angle=-90, width=1.00\textwidth]{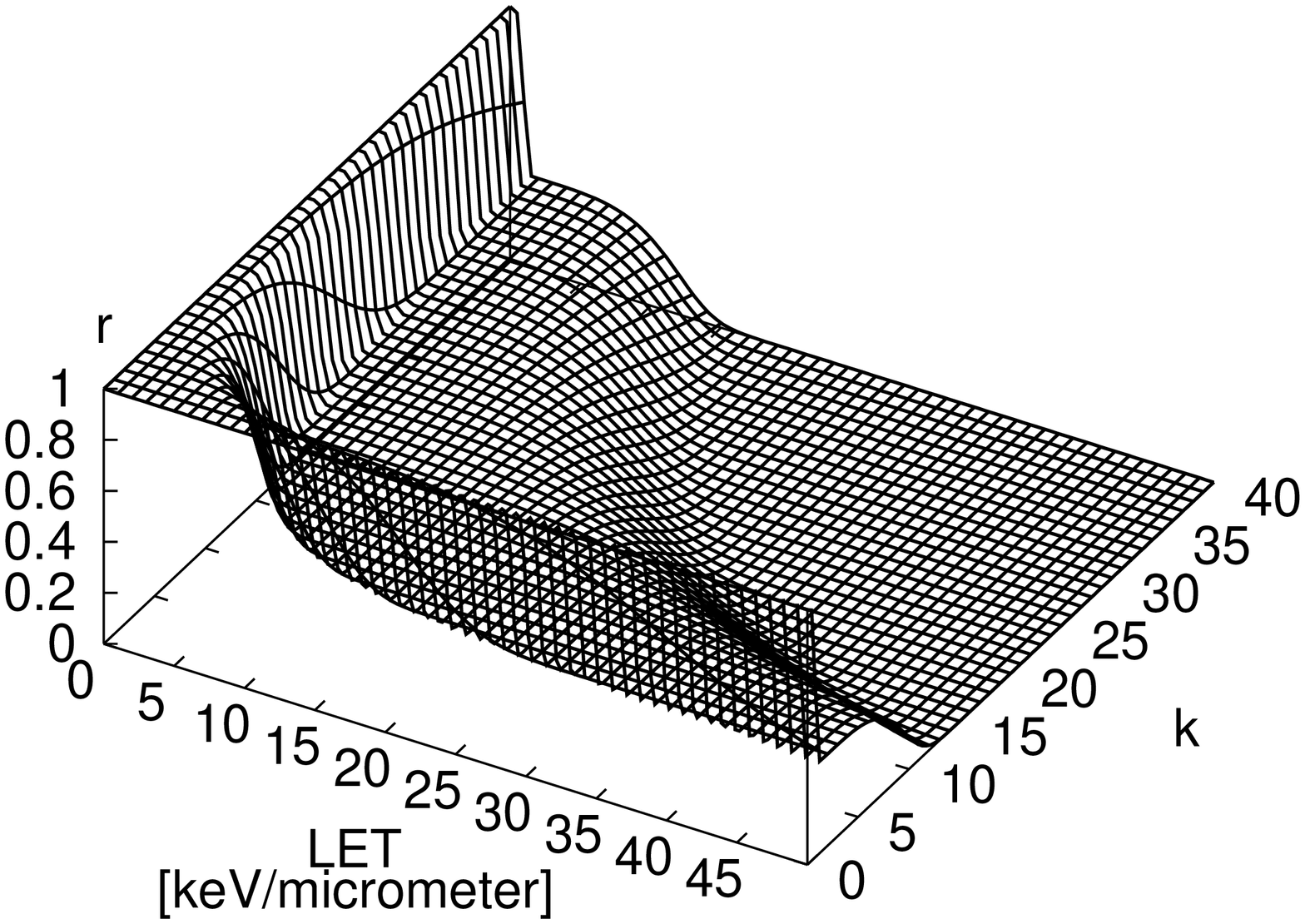}
	\end{minipage}
	\caption{Left panel: The probabilities of single-particle induced, $a$ (solid line), and pair DNA damages, $b$, as functions of proton LET values. Right panel: The probability of successful repair, decreasing with both LET and number of traversing protons, reflecting the increasing complexity of DNA damage.}
	\label{fig:proton-abr}
\end{figure}

Model calculations of the survival curves (full lines) together with experimental data (points) are shown in Fig.~\ref{fig:proton}. For comparison, fits of the data according to the linear-quadratic (LQ) model are displayed, too. Probabilities of forming "single-particle" and "pair" DNA damages discussed in the preceding are shown in Fig.~\ref{fig:proton-abr}. Single-particle induced damages, $a$, have turned out to play a minor role only. Consequently, it has been possible to neglect their repair, i.e.\ to put $r^A(k,\lambda) \equiv 0$, which has simplified the fitting procedure. The derived probability of successful repair of the "pair damages" as a function of traversing particle numbers $k$ and LET values $\lambda$ is plotted in Fig.~\ref{fig:proton-abr}.

The optimization procedure used in the present analysis has been based on applying standard minimization methods; mainly the SIMPLEX and MIGRAD methods implemented in the MINUIT minimization tool~\citep{MINUIT} have been applied to dedicated computer codes written in the programming language FORTRAN. The goodness of fit has been assessed by calculating the $\chi^2$ values~\citep{Groom} for individual survival curves. They have been calculated as a sum of the deviations of (logarithms of) model calculations, $s_{model}$, from the data, $s_{exp}$, weighted by the experimental errors, $\Delta_i$, of individual data points (again in the logarithmic scale),
\[ \chi^2 = \sum_i \frac{(\ln s_{model} - \ln s_{exp})^2} {\Delta_i^2}  \ . \]
The values of $\chi^2$ for individual survival curves represented by the probabilistic two-stage (P2S) model and by the linear-quadratic (LQ) model are listed in Table~\ref{tab:chi2}. The total $\chi^2$ values were 24.0 for the P2S model and 45.0  for the LQ model, respectively.
%(LET values from 7.7 to 37.8~$\mathrm{keV/\mu m}$) were 2.78, 3.84, 0.704, 9.53, 4.44, and 2.73 (total $\chi^2 = 24.0$). For comparison, $\chi^2$ values obtained by fitting the data by the LQ model were 2.98, 4.36, 1.08, 6.06, 14.4, and 15.8, respectively (total $\chi^2 = 45.0$).
The corresponding values of auxiliary parameters involved in Eqs.~(\ref{eq:proton}) are listed in Table~\ref{tab:parameters}. These values correspond to the effective cross-section of V79 nucleus (or chromosomal system) $\sigma = 12.8 \ \mathrm{\mu m^2}$; the goodness of fit of the model to the data has varied only slightly around this value. 
%\textbf{list/table}

%%%%%%%%%%%%%%%%%%%%%%%%%%%%%%%%%%%%%%%%%%%%%%%%%%%%%%%%%%%%%%%%%%%

\begin{table}
\caption{\label{tab:chi2}Experimental survival curves and precision of representing them by the probabilistic two-stage (P2S) and the linear-quadratic (LQ) models.}
\lineup
\begin{indented}
\item[]\begin{tabular}{@{}lllllll}
\br
%
% TO ALIGN NUMBERS AT DECIMAL POINT, use the commented version:
%
%Energy [MeV]&5.01&\03.20&\01.41&\00.76&\00.64&\00.57\\
%LET [$\mathrm{keV/\mu m}$]&7.7&11.0&20.0&30.5&34.6&37.8\\
%\mr
%$\chi^2_{\mathrm{P2S}}$&2.78&\03.84&\00.70&\09.53&\04.44&\02.73\\
%$\chi^2_{\mathrm{LQ}}$&2.98&\04.36&\01.08&\06.06&14.4&15.8\\
%
% NOT TO ALIGN NUMBERS AT DECIMAL POINT, use this one:
%
Energy [MeV]&5.01&3.20&1.41&0.76&0.64&0.57\\
LET [$\mathrm{keV/\mu m}$]&7.7&11.0&20.0&30.5&34.6&37.8\\
\mr
$\chi^2_{\mathrm{P2S}}$&2.78&3.84&0.70&9.53&4.44&2.73\\
$\chi^2_{\mathrm{LQ}}$&2.98&4.36&1.08&6.06&14.4&15.8\\
\br
\end{tabular}
\end{indented}
\end{table}

\begin{table}
\caption{\label{tab:parameters}Values of auxiliary parameters involved in model representation of damage induction and repair probabilities, Eq.~(\ref{eq:proton}).}
\begin{indented}
\item[]\begin{tabular}{@{}lll}
\br
Single-particle induced damage&Pair damage&Repair probability\\
$a(\lambda)$&$b(\lambda)$&$r^B(k,\lambda)$\\
\mr
$a_1=0.0022 \ \mathrm{\mu m/keV}$&$b_1=0.12 \ \mathrm{\mu m/keV}$&$r_1=0.024 \ \mathrm{\mu m/keV}$\\
$a_2=0.013 \ \mathrm{(\mu m/keV)^2}$&$b_2=5.0$&$r_2=5.0$\\
$a_3=0.026 \ \mathrm{\mu m/keV}$&$b_3=24.1$&$r_3=0.56$\\
$a_4=5.0$&$b_4=0.061 \ \mathrm{\mu m/keV}$&$r_4=0.0021 \ \mathrm{\mu m/keV}$\\
&$b_5=1.76$&$r_5=5.0$\\
\br
\end{tabular}
\end{indented}
\end{table}

%%%%%%%%%%%%%%%%%%%%%%%%%%%%%%%%%%%%%%%%%%%%%%%%%%%%%%%%%%%%%%%%%%%%%%%%%%%%%%%%
\section{Discusion}
\label{sec:Discusion}

Basic characteristics of DNA damage formation and of cellular repair processes have been derived in the studied case. The corresponding probabilities have been determined by analyzing experimental cell survival data only; no other inputs or assumptions have been used. Nevertheless, the results are consistent with experimental studies concerning the yields of DNA damage; compare e.g.~\citep{Prise-DNA}.

The results have shown that for protons in the studied range of rather high LET, the pair damages dominate over the single-particle ones. In the local-effect model (LEM), for example, combined effects are not taken into account at all, as all damages are assumed to be of lethal nature. This seems to be the reason why discrepancies of the order of 25\% have been found between the given experimental data and the corresponding calculations based on the LEM scheme; compare~\citep{Belli, Scholz+Kraft-1995}. The differences in inactivation mechanism between protons and heavier ions, identified on the basis of the given detailed model scheme, will be discussed in detail in a forthcoming paper~\citep{protons-ions}.

The formation of lethal damages by individual particles (single-particle induced damages) gets saturated for protons at LET values around 30~$\mathrm{keV/\mu m}$. This means that for protons the RBE effects related to overkill occur already on the level of damage induction probability, which has a maximum at values much lower than unity (around 20\% in the present analysis). The probability of inducing combined damages, on the other hand, possesses a monotonous increase over the whole studied LET range. The repair probability exhibits decreasing behavior with increasing LET and/or particle numbers, corresponding to increasing complexity of the total damage induced in chromosomal DNA.

The effective cross-section of cell nuclei has been established in the present analysis on a phenomenological basis only. It has been limited from below by the given data, as by assuming a too low value of the cross section the cell survival probability would be too high as compared to the given data. Upper limit on the effective cross section has been represented by the average geometrical cross-section of cell nuclei of the given cell line $\sigma_{geom} = 134 \, \mathrm{keV/\mu m}$ reported in the given experiment~\citep{Belli}. The effective cross-section (i.e., the cross-section of the sensitive region within the nucleus with respect to proton tracks) has corresponded to approximately 10\% of the nucleus.

In the future, attempts will be made to relate the damage induction and repair probabilities to the results of track structure studies and other microscopic models of physical and biological processes. These models might be helpful in determining the effective cross section of the cell nuclei with respect to a given ion (and its energy), too. The present analyses rely on the fact that the most efficient parts of proton tracks are much narrower than the dimensions of cell nuclei, which has enabled to apply Poisson statistics.

As to the precision of representing experimental cell-survival data, the present calculations and those based on the LQ model yield similar results for lower-LET protons, where the experimental survival curves possess quite simple shapes. They differ significantly in the highest LET values (compare Fig.~\ref{fig:proton}) corresponding to proton track ends, where detailed structure of the data can be represented only by the detailed scheme of the probabilistic two-stage model.

The required level of accuracy influences the number of auxiliary parameters that have to be involved in the parameterization of DNA damage induction and repair probabilities. If e.g.\ in treatment planning applications a less precise representation of survival curves is sufficient, less detailed parameterization e.g.\ of the form of $f(x) = \exp(-(\alpha x)^\beta)$ might be used for $b(\lambda)$ and $r^B(k,\lambda)$ instead of those of Eqs.~(\ref{eq:proton}). This difference in parameterization is of technical nature only, the basic scheme of the model remaining the same.

%%%%%%%%%%%%%%%%%%%%%%%%%%%%%%%%%%%%%%%%%%%%%%%%%%%%%%%%%%%%%%%%%%%%%%%%%%%%%%
\section{Conclusion}
\label{sec:Conclusion}
The probabilistic two-stage model provides a realistic description of the radiobiological mechanism. It includes description of DNA damages of different severity caused by individual traversing particles, as well as of their synergetic and saturation combinations. Basic characteristic of repair processes are incorporated, too. The model might serve as a basis for more detailed microscopic modeling of radiobiological effects.

As demonstrated explicitly for the case of low-energy protons, the probabilistic two-stage model enables to represent experimental cell survival curves with much better precision than the LQ model, while the total number of parameters involved is about the same. The ability of representing cell survival precisely is a prerequisite if different fractionation schemes of hadron radiotherapy are to be evaluated, because in fractionated irradiation the local variations (i.e.\ local deviations of the cell survival ratio from the global shape of survival curve) may get largely amplified in the cumulative effect of the given scheme. The model enables to represent simple linear and parabolic, but also rather complex survival curves, including e.g.\ the low-dose hypersensitivity phenomenon, where a kind of triggering behavior of repair probability has to be taken into account. Unconventional fractionation schemes including hyperfractionation might be assessed on this model basis, too.

%%%%%%%%%%%%%%%%%%%%%%%%%%%%%%%%%%%%%%%%%%%%%%%%%%%%%%%%%%%%%%%%%%%%%%%%

%\begin{thebibliography}{999}%{breitestes Label neboli uplne nejdelsi klic jaky tam muze byt}%{999}%{breitestes Label}
\References

	\bibitem[Belli \etal(1997)]{Belli-Semiempirical}
	Belli M, Campa A, Ermolli I 1997
	A semi-empirical approach to the evaluation of the relative biological effectiveness of therapeutic proton beams: The methodological framework
	\emph{Radiat. Res.} \textbf{148} 592--598

	\bibitem[Belli \etal(1998)]{Belli}
%	Belli M., Cera F., Cherubini R. et al (1998):
	Belli M, Cera F, Cherubini R, Dalla Vecchia M, Haque AMI, Ianzini F, Moschini G, Sapora O, Simone G, Tabocchini MA, Tiveron P 1998
	RBE-LET relationships for cell inactivation and mutation induced by low energy protons in V79 cells: further results at the LNL facility.
	\emph{Int. J. Radiat. Biol.} \textbf{74} 501--509

	\bibitem[Belli \etal(2000)]{Belli 2000}
%	Belli M., Bettega D., Calzolari P. et al (2000):
	Belli M, Bettega D, Calzolari P, Cera F, Cherubini R, Dalla Vecchia M, Durante M, Favaretto S, Gialanella G, Grossi G, Marchesini R, Moschini G, Piazzola A, Poli G, Pugliese M, Sapora O, Scampoli P, Simone G, Sorrentino E, Tabocchini MA, Tallone L, Tiveron P 2000
	Inactivation of human normal and tumour cells irradiated with low energy protons
	\emph{Int. J. Radiat. Biol.} \textbf{76} 831--839

	\bibitem[Chalmers \etal(2004)]{Chalmers}
	Chalmers A, Johnston P, Woodcock M, Joiner M, Marples B 2004
	PARP-1, PARP-2, and the cellular response to low doses of ionizing radiation
	\emph{Int. J. Radiat. Oncol. Biol. Phys.} \textbf{58} 410--419

	\bibitem[Folkard \etal(1996)]{Folkard}
	Folkard M, Prise K M, Vojnovic B, Newman H C, Roper M J, Michael B D 1996
	Inactivation of V79 cells by low-energy protons, deuterons and helium-3 ions
	\emph{Int. J. Radiat. Biol.} \textbf{69} 729--738

	\bibitem[Hagiwara \etal(2002)]{Groom}
%	Groom et al. 2000 - Particle Data Group
%	Groom et al.: Particle Data Group, http://pdg.lbl.gov/pdg.html \\
%	Groom D.E. et al: 2000 Review of Particle Physics. \emph{The European Physical Journal} C15, 1, 2000.
	Hagiwara K, Hikasa K, Nakamura K \etal 2002
	Review of Particle Physics
	\emph{Phys. Rev.} D\textbf{66} 010001

	\bibitem[Holley \etal(1990)]{Holley}
	Holley WR, Chatterjee A, Magee JL 1990
	Production of DNA strand breaks by direct effects of heavy charged-particles
	\emph{Radiat. Res.} \textbf{121} 161--168

	\bibitem[James(1994)]{MINUIT}
	James F 1994
	MINUIT Minimization package - Reference Manual
	\emph{CERN Program Library Long Writeup} D506, CERN Geneva

	\bibitem[Joiner(1997)]{Steel-Joiner}
	Joiner MC 1997
	Models of radiation cell killing.
	In: \emph{Basic clinical radiobiology}, Ed. Steel GG, Arnold, London, pp 52--57

	\bibitem[Joiner \etal(2001)]{Joiner}
	Joiner MC, Marples B, Lambin P, Short SC, Turesson I 2001
	Low-dose hypersensitivity: current status and possible mechanisms
	\emph{Int. J. Radiat. Oncol. Biol. Phys.} \textbf{49} 379--389

	\bibitem[Judas and Lokaj\'{\i}\v{c}ek(2001)]{Judas+Lok-JTBi}
	Judas L, Lokaj\'{\i}\v{c}ek M 2001
	Cell inactivation by ionizing particles and the shapes of survival curves
  \emph{J. Theor. Biol.} \textbf{210} 15--21

	\bibitem[Kanai \etal(1999)]{HIMAC}
	Kanai T, Endo M, Minohara S, Miyahara N, Koyama-Ito H, Tomura H, Matsufuji N, Futami Y, Fukumura A, Hiraoka T, Furusawa Y, Ando K, Suzuki M, Soga F, Kawachi K 1999
%	T. Kanai, M. Endo, S. Minohara, N. Miyahara, H. Koyama-Ito, H. Tomura, N. Matsufuji, Y. Futami, A. Fukumura, 	K. Kawachi, (1999)
	Biophysical characteristics of HIMAC clinical irradiation system for heavy-ion radiation therapy 
	\emph{Int. J. Radiat. Oncol. Biol. Phys.} \textbf{44} 201--210

	\bibitem[Kraft \etal(1997)]{Kraft-WhatKindOfRBShouldBeDone}
	Kraft G, Kraft-Weyrather W, Taucher-Scholz G, Scholz M 1997
	What kind of radiobiology should be done at a hadrontherapy centre?
	In: \emph{Advances in Hadrontherapy}, 
	Ed. Amaldi U, Larsson B, Lemoigne Y, Elsevier, pp 38--54

	\bibitem[Kraft \etal(1999)]{Scholz 1999}
	Kraft G, Scholz M, Bechthold U 1999
	Tumor therapy and track structure
	\emph{Radiat. Environ. Biophys.} \textbf{38} 229--237

	\bibitem[Kundr\'{a}t(2004)]{PhD}
	Kundr\'{a}t P 2004
	Mechanism of biological effects of accelerated protons and light ions and its modelling
	\emph{PhD thesis} Charles University in Prague

	\bibitem[Kundr\'{a}t \etal(2004)]{protons-ions}
	Kundr\'{a}t P, Lokaj\'{\i}\v{c}ek M, Hrom\v{c}\'{\i}kov\'{a} H 2004
	On the qualitative difference between the cell-inactivation mechanism of protons and light ions.
	Paper being prepared for publication.

	\bibitem[Miller \etal(1999)]{Brenner}%Brenner - single alpha particles
	Miller RC, Randers-Pehrson G, Geard CR, Hall EJ, Brenner DJ 1999
	The oncogenic transforming potential of the passage of single alpha particles through mammalian cell nuclei
	\emph{Proc. Natl. Acad. Sci. U.S.A.} \textbf{96} 19--22

	\bibitem[Niemierko \etal(1992)]{Niemierko+Urie+Goitein}
	Niemierko A, Urie M, Goitein M 1992
	Optimization of 3D radiation-therapy with both physical and biological end-points and constraints
	\emph{Int. J. Radiat. Oncol. Biol. Phys.} \textbf{23} 99--108

	\bibitem[Niemierko and Goitein(1993)]{Niemierko+Goitein}
	Niemierko A, Goitein M 1993
	Modeling of normal tissue-response to radiation -- the critical volume model
	\emph{Int. J. Radiat. Oncol. Biol. Phys.} \textbf{25} 135--145

	\bibitem[Prise \etal(1998)]{Prise-DNA}
	Prise KM, Ahnstrom G, Belli M, Carlsson J, Frankenberg D, Kiefer J, Lobrich M, Michael BD, Nygren J, Simone G, Stenerlow B 1998
%	K. M. Prise, G. Ahnstrom, M. Belli, J. Carlsson, D. Frankenberg, J. Kiefer, M. Lobrich, B. D. Michael, J. Nygren, B. Stenerlow, 
	A review of DSB induction data for varying quality radiations
	\emph{Int. J. Radiat. Biol.} \textbf{74} 173--184

	\bibitem[Schettino \etal(2001)]{Schettino}
	Schettino G, Folkard M, Prise KM, Vojnovic B, Bowey AG, Michael BD 2001
%	Schettino G., Folkard M., Prise K.M. et al:
	Low-dose hypersensitivity in Chinese hamster V79 cells targeted with counted protons using a charged-particle microbeam
	\emph{Radiat. Res.} \textbf{156} 526--534

	\bibitem[Scholz and Kraft(1994)]{LEM-1}
	Scholz M, Kraft G 1994
	Calculation of heavy-ion inactivation probabilities based on track structure, x-ray-sensitivity and target size
	\emph{Radiat. Prot. Dosimetry} \textbf{52} 29--33

	\bibitem[Scholz and Kraft(1995)]{Scholz+Kraft-1995}
	Scholz M, Kraft G 1995
	Track structure and the calculation of biological effects of heavy charged-particles
	\emph{Adv. Space Res.} \textbf{18} 5--14

	\bibitem[Scholz \etal(1997)]{LEM-2}
	Scholz M, Kellerer AM, Kraft-Weyrather W, Kraft G 1997
	Computation of cell survival in heavy ion beams for therapy -- The model and its approximation
	\emph{Radiat. Environ. Biophys.} \textbf{36} 59--66

	\bibitem[Webb and Nahum(1993)]{Webb+Nahum}
	Webb S, Nahum AE 1993
	A model for calculating tumour control probability in radiotherapy including the effects of inhomogeneous distributions of dose and clonogenic cell density
	\emph{Phys. Med. Biol.} \textbf{38} 653--666

	\bibitem[Weyrather \etal(1999)]{Weyrather}
	Weyrather WK, Ritter S, Scholz M, Kraft G 1999
	RBE for carbon track-segment irradiation in cell lines of differing repair capacity
	\emph{Int. J. Radiat. Biol.} \textbf{75} 1357--1364

\endrefs
%\end{thebibliography}	

\end{document}